\documentclass[aps,prd,amssymb,showpacs,floatfix,preprint,superscriptaddress,nofootinbib]{revtex4-1}
\usepackage{amsmath,amsfonts,graphics,epsfig,amssymb,color}

\newcommand{\blue}{\protect\color{blue}}

\begin{document}
\title{Linking infrared and ultraviolet parameters of pion-like states in
strongly coupled gauge theories}

\author{S.V.~Troitsky}
\affiliation{Institute for Nuclear
Research of the Russian Academy of Sciences, 60th October Anniversary
Prospect 7a, Moscow 117312, Russia}
\email{st@ms2.inr.ac.ru}

\author{V.E.~Troitsky}
\affiliation{D.V.~Skobeltsyn Institute of Nuclear Physics,
M.V.~Lomonosov Moscow State University, Moscow 119991, Russia}
\email{troitsky@theory.sinp.msu.ru}



\begin{center}
\begin{abstract}
It has been shown previously that in a relativistic
constituent-quark model, predictions for the electromagnetic form factor
of the $\pi$ meson match not only experimental data but also, in the limit
of large momentum transfers, the asymptotics derived from Quantum
Chromodynamics (QCD). This is remarkable since no parameters are
introduced to provide for this infrared-ultraviolet link. Here, we follow
this approach, going beyond QCD. We obtain numerical relations between the
gauge coupling constant, the decay constant and the charge radius of the
pion-like meson in general strongly-coupled theories. These relations are
compared to published lattice results for $SU(2)$ gauge theory with two
fermion flavours, and a good agreement is demonstrated. Further
applications of the approach, to be explored elsewhere, include composite
Higgs and dark-matter models.
\end{abstract}
\end{center}
\maketitle

\section{Introduction}
\label{sec:intro}
Obtaining first-principle quantitative predictions concerning stroungly
coupled bound states remains the main challenege of quantum field theory.
The only available direct method, lattice calculations, is complicated and
resource consuming in practical implementation, especially when light
fermions are involved. Numerous semi-phenomenological approaches have been
put forward in order to obtain quantitative description of bound states in
Quantum Chromodynamics (QCD), mesons and baryons. In many cases,
underlying symmetries of the theory were used as guiding principles in
these calculations.

One of these approaches, based on Dirac~\cite{Dirac} Instant Form of the
Relativistic Hamiltonian Dynamics (RHD; for reviews, see
Refs.~\cite{LeS78, KeisterPolyzou, Coe92}), has been particularly
successful in description of electromagnetic properties of light mesons. A
distinctive feature of this approach is that the Poincar\'e invariance is
fully exploited and kept unbroken at all steps. The canonical example of
the approach's application was the calculation of the $\pi$-meson
electromagnetic form factor, $F_{\pi}$, as a function of the momentum
transfer, $Q^{2}$. It has been shown that quantitative results for
$F_{\pi}(Q^{2})$ are robust with respect to variations in the most
uncertain ingredient of the approach, the phenomenological wave function
$\phi(k)$, provided the pion decay constant, $f_{\pi}$, is
fixed~\cite{KrTr-EurPhysJ2001-bundles}. Numerical results therefore depend
on two phenomenological parameters, one combination of which is fixed
through $f_{\pi}$. The remaining combination was fitted (in 1998) from the
condition that $\left. F_{\pi}(Q^{2}) \right|_{Q^{2}\to 0}$ reproduced
correctly the experimental data points~\cite{Amendolia}, that is, the pion
charge radius~\cite{KrTr-EurPhysJ2001-bundles}. Subsequent
measurements~\cite{exp-data, Horn} of $F_{\pi}(Q^{2})$ spanning an order
of magnitude larger values of $Q^{2}$ agreed with the prediction of the
model surprisingly well~\cite{KrTr-PRC2009}. In addition, it has been
demonstrated that the very same model predicts also the correct QCD
asymptotics~\cite{Qcounting1, Qcounting2, FarrarJackson,
EfremovRadyushkinAs, LepageBrodskyAs} of $F_{\pi}(Q^{2})$ at large
$Q^{2}$, reproducing both the functional dependence~\cite{KrTr-asymp} on
$Q^{2}$ and, for the very same choice of parameters, the numerical
coefficient~\cite{TrTr-PRD2013-asympt}. This is achieved when the
constituent-quark mass is switched off~\cite{Kiss}, independendently of
the way of this switching.

Altogether, these successes look unusual and may indicate that the model
catches some basic dynamical features of the $\pi$ meson thanks to the full
incorporation of the relativistic invariance. This might open the
possibility to go beyond QCD and to apply the model to pion-like bound
states in other hypothetical strongly coupled theories, e.g.\ those
describing composite Brout-Englert-Higgs scalar~\cite{0911.0931,
1402.0233, 1502.04718} and/or composite dark-matter
particles~\cite{1703.06903}. At the same time this invites further
quantitative tests of the approach which, given the lack of well
established strongly coupled gauge theories in Nature, may be performed
only by comparison with lattice results or limiting cases. This is the
subject of the present work.

The rest of the paper is organized as follows. In Section~\ref{sec:motiv},
we give a very brief account of the model and refer to previous works
where all details can be found. We discuss in more detail manifestations
of the success of the model which motivate the present study.
Section~\ref{sec:main} presents the method allowing to relate
quantitatively parameters of the gauge theory and low-energy meson
properties. In Section~\ref{sec:lat}, we address a few examples of non-QCD
gauge theories for which lattice calculation of the form factor of a
pion-like state has been reported, and compare lattice results with those
obtained within our approach. We briefly conclude and discuss future
applications of our method in Section~\ref{sec:concl}.

\section{Motivation}
\label{sec:motiv}
The model we discuss here~\cite{KrTr-JHEP, long, KrTr-PRC2002,
KrTr-PRC2003, EChAYa2009} has been developed for the description of
electroweak properties of light strongly interacting two-particle bound
states and has been succesfully applied to the
deuteron~\cite{KrTrTs-PRC2008-Deuteron},
$\pi$~\cite{KrTr-EurPhysJ2001-bundles, KrTr-PRC2009},
$\rho$~\cite{KrPolTr-PRD2016-RHOmeson, KrPolTr-PRD2018-RHOmeson} and
$K$~\cite{KrTr-EPJC2017-Kmeson} mesons. The model is based on the instant
form of the relativistic Hamiltonian dynamics (see e.g.\
Ref.~\cite{KeisterPolyzou}), supplemented by the so-called modified
impulse approximation~\cite{KrTr-PRC2002}, which is the key ingredient of
the approach since it removes certain disadvantages of the instant form.
The form factors can be obtained with the use of the Wigner-Eckart theorem
for the Poincar\'e group~\cite{KrTr-TMF2005}. Here, we will focus on the
bound states similar to the charged $\pi$ meson, for which all details and
explicit formulae are given in Refs.~\cite{KrTr-EurPhysJ2001-bundles,
KrTr-PRC2009} (see the Appendix of Ref.~\cite{TrTr-PRD2013-asympt} for a
useful summary). The essential feature of the method, which distinguishes
it from many other approaches (see e.g.\ Ref.~\cite{HornRoberts-pion} for
a
recent review), is that the explicit Poincar\'e invariance is kept
throughout the calculation.

The model of the electromagnetic structure of the $\pi$ meson has two
principal phenomenological parameters, the constituent-quark mass $M$ and
the meson wave-function scale $b$. The latter is a dimensionful parameter
whose definition depends on the particular choice of the wave function;
however, it has been shown in Ref.~\cite{KrTr-EurPhysJ2001-bundles} that
the dependence on the shape of the wave function diminishes provided the
pion decay constant, $f_{\pi}$, is fixed (like other observable
quantities, it is expressed through $M$ and $b$, see
Section~\ref{sec:main}).
In Ref.~\cite{KrTr-EurPhysJ2001-bundles}, the dependence of the form
factor on the choice of the wave function was studied. Three different
wave-function shapes were considered and it has been shown that the
dependence on the shape of the wave function diminishes provided the pion
decay constant, $f_\pi$, is fixed (like other observable quantities, it is
expressed through M and b, see Section~\ref{sec:main}). At large $Q^2$, the
variation in the value of $F_\pi$ at fixed $M$ and $f_\pi$ with the change
of the wave function does not exceed $\sim 3\%$ for most values of
parameters, though it becomes larger at very large values of $M \gg
f_{\pi}$. We estimate the related systematic uncertainty of the method as
$\sim 5\%$ and, following previous studies~\cite{TrTr-PRD2013-asympt}, use
the power-law type wave function~\cite{wave-function} in the momentum
($k$) space,
\begin{equation}
\phi(k)\propto \left(4\left(k^{2}+M^{2} \right)
\right)^{1/4}\,k\,\left(k^2/b^{2}+1 \right)^{-3},
\label{Eq:5*}
\end{equation}
for our numerical examples.

In principle, the model allows for inclusion of two other parameters whcih
have minor impact on numerical results and were never varied; they are
related to deviations from point-like constituent quarks and affect the
form factor at high mometum transfers, $Q^{2}$. One is the coefficient $C$
in the relation between the constituent-quark mass $M$ and its effective
radius, $C/M$; it was always fixed at $C=0.3$ in previous works and so we
do here. The impact of the second parameter, the sum $s_{q}$ of the
anomalous magnetic moments of quarks, is clearly within the overall
uncertainty of the method for hypothetical non-QCD theories.
For QCD, $s_q$ was determined with the help of Gerasimov sum
rules~\cite{Gerasimov:1996ci} and was found to be $s_q \approx 0.03$; it
enters the expressions for the form factor in a sum with quark and
antiquark charges (equal to $1\gg 0.03$) and is therefore expected to have
a minor impact on the result. Indeed, we have checked numerically that the
effect of its variation within $0 \le s_q \le 0.1$ on the form-factor
asymptotics is negligible compared to other uncertainties of the model.
Similar sum rules justifying a particular value of $s_q$ are unavailable
for a general non-QCD theory, and in the numerical calculations presented
here we simply put $s_q = 0$.

We turn now to the motivations behind the extension of the $\pi$-meson
model to a general strongly coupled theory, which we propose and start to
study here.

\begin{figure}[tbp]
\centering
\includegraphics[width=.85\textwidth%
]{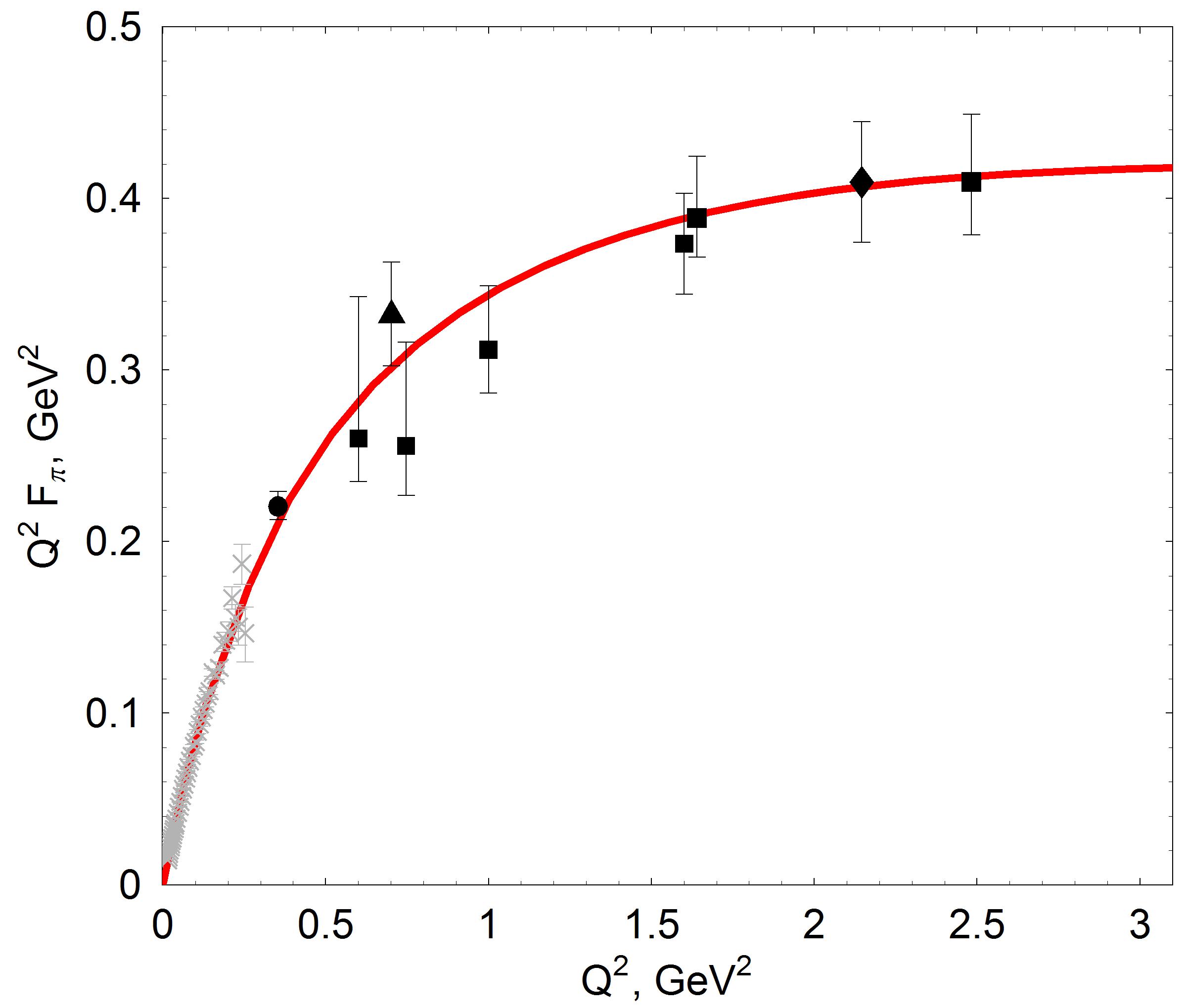}
\caption{\label{fig:data} Motivation~I. The $\pi$ meson form factor:
model predictions (full line) versus experimental data. Black data points
have been published after the prediction, while gray data points at
$Q^{2}\to 0$ were used to make the prediction. Symbols correspond to
different data sets: Refs.~\cite{Amendolia} (crosses), \cite{Ackerman}
(reanalized in Ref.~\cite{exp-data}, circle), \cite{Brauel} (reanalized in
Ref.~\cite{exp-data}, triangle), \cite{Horn} (diamond) and \cite{exp-data}
(squares).}
\end{figure}

The first motivation is the predictivity of the model. In 1998, the two
parameters, $M$ and $b$, were fixed by fitting two observable quantities,
the decay constant, $f_{\pi}$, and the charge radius,
$\langle r_{\pi}^{2} \rangle ^{1/2}$, of the $\pi$ meson. This made it
possible to calculate the form factor, $F_{\pi}(Q^{2})$, as the function
of the momentum transfer. Figure~\ref{fig:data}
presents the predicted function $F_{\pi}(Q^{2})$ together with
experimental data points: data shown in gray were obtained earlier and
were used in the fit through $\langle r_{\pi}^{2} \rangle ^{1/2}$, while
black data points, spanning a further order of magnitude in $Q^{2}$, have
been obtained after the prediction. The new data have demonstrated an
impressive agreement with the calculation ($\chi^{2}\approx 4.4$ for 9
degrees of freedom, no free parameter).

\begin{figure}[tbp]
\centering
\includegraphics[width=.85\textwidth%
]{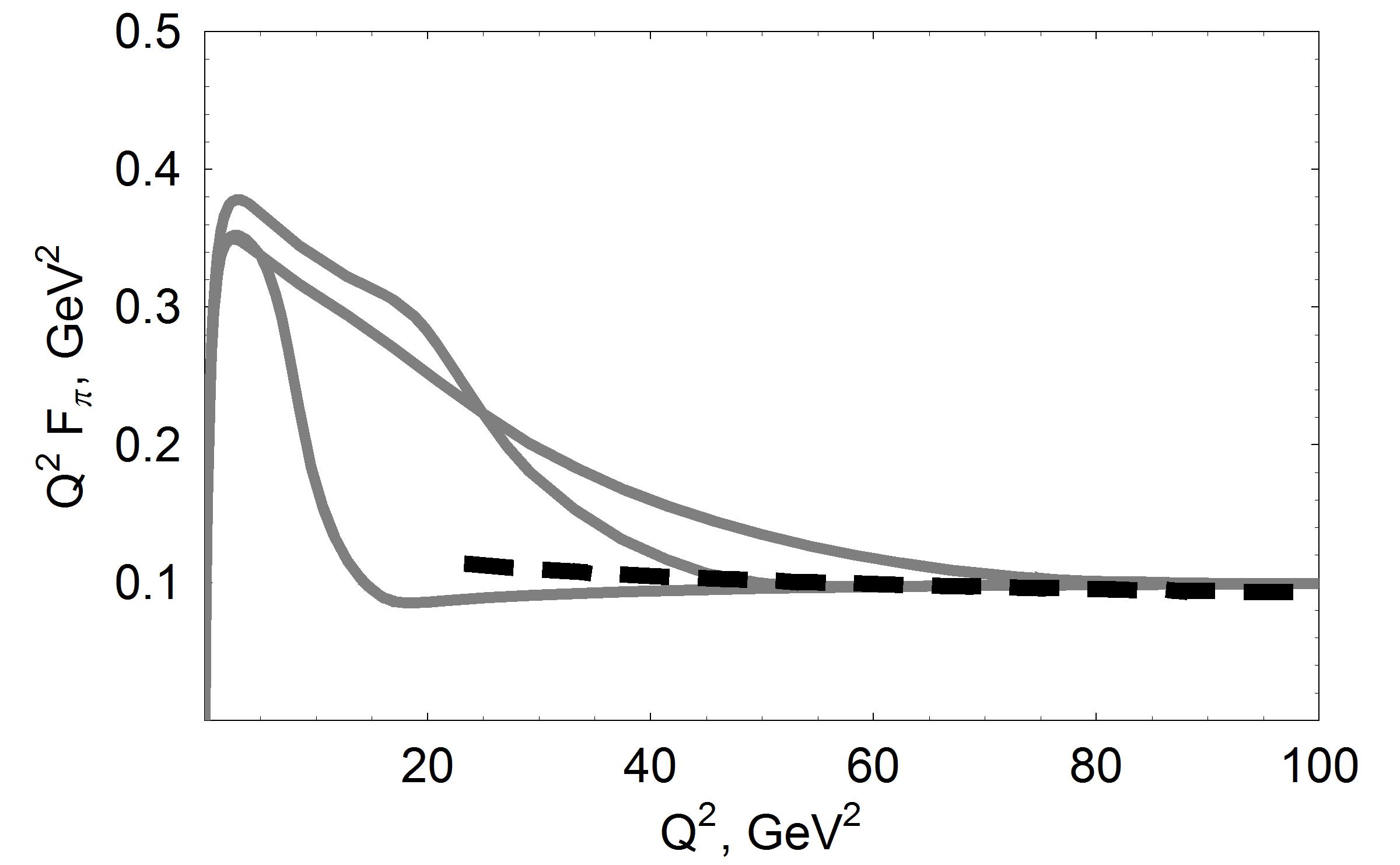}
\caption{\label{fig:as} Motivation~II. Once the constituent-quark mass is
switched off, $F_{\pi}(Q^{2})$ reaches the asymptotics predicted from
QCD~\cite{FarrarJackson, EfremovRadyushkinAs, LepageBrodskyAs}, shown as
the thick dashed line. Three thin curves correspond to different ways of
switching $M$ off (the explicit expressions of $M(Q^2)$, inspired by
Ref.~\cite{Kiss}, are given in Ref.~\cite{TrTr-PRD2013-asympt}). The
asymptotics does not depend on the details how $M$ is set to zero but does
depend on its infrared value.}
\end{figure}

The second motivation is the possibility to relate infrared and
ultraviolet physics within a single model. In Ref.~\cite{KrTr-asymp}, it
has been shown that the model reproduces the \emph{functional form} of the
QCD asymptotics~\cite{Qcounting1, Qcounting2} for $F_{\pi} (Q^{2})$
provided the constituent-quark mass is switched off, $M\to 0$, at $Q^{2}\to
\infty$. Moreover, numerical calculations of
Ref.~~\cite{TrTr-PRD2013-asympt} have demonstrated that the
\emph{coefficient} of the asymptotics~\cite{FarrarJackson,
EfremovRadyushkinAs, LepageBrodskyAs} is also reproduced correctly. This
is achieved independently of the way $M$ is switched off (see
Figure~\ref{fig:as}),
and therefore the correct ultraviolet asymptotics of $F_{\pi}(Q^{2})$ is
reproduced without any additional parameters (with respect to those two
present in the successful infrared model). Reaching the QCD asymptotics is
an infrequent, though welcome, feature of infrared models of mesons; to
our best knowledge, no other model achieves it without introduction of new
tunable parameters. For the QCD case, unusual predictivity of the model
allowed for construction of a ``parameter-free model of electromagnetic
properties of light mesons'', successfuly connecting corresponding
observables of $\pi$, $\rho$ and $K$
mesons~\cite{KrPolTr-PRD2016-RHOmeson, KrPolTr-PRD2018-RHOmeson,
KrTr-EPJC2017-Kmeson}.

Meanwhile, methods to relate infrared and
ultraviolet physics of strongly coupled gauge theories are in demand for
numerous extensions of the Standard Model, notably including models with
composite Brout-Englert-Higgs scalars (see e.g.\ Refs.~\cite{0911.0931,
1402.0233, 1502.04718} and references therein) and/or dark-matter
particles (in particular, within the asymmetric dark-matter scenario
\cite{asym-1, asym-2}; see e.g.\ Ref.~\cite{1703.06903} for a recent
discussion and collection of references). Recently, the interest in models
with new strong dynamics increased considerably because of their possible
relation to flavour anomalies observed at the Large Hadron Collider (LHC)
\cite{1704.05438, 1706.07808}. In the context of our study, it is
interesting to note that in certain scenarios, electromagnetic form
factors of composite dark-matter particles determine their cross sections
\cite{0812.3406} and their knowledge is therefore crucial for tests of the
viability of the models and of their experimental and observational
predicitons.

Theoretical approaches to calculation of these observables are limited to
lattice gauge theories whose interpretation in terms of the underlying
continuum renormalizable theory is not always straightforward (see
Section~\ref{sec:lat}). Another problem for practical applications of the
lattice approach is that the numerical methods required to reach
reasonable precision in description of light fermion bound states consume
lots of computational resources. It is therefore tempting to generalize
our method to non-QCD theories, which would open a possibility of fast
quantitative description of light bound states (at least of their
electromagnetic properties). But how can one verify that the method works
and that both successes (prediction of the measured form factor and
parameter-free matching with the QCD asymptotics) were not just lucky
coincidences? In the absence of a first-principle derivation of our
results from a gauge field theory, we can nevertheless test it by
comparison with particular lattice results, see Section~\ref{sec:lat}.

\section{Relating the form factor, the meson decay constant and the
gauge coupling constant.}
\label{sec:main}
Consider a QCD-like strongly coupled gauge theory allowing for a pion-like
state, $\Pi$ (we keep the notion $\pi$ for the QCD $\pi$ meson).
``Pion-like'' means that $\Pi$ is a meson (a bound state of a fundamental
$q_{i}$ and antifundamental $\bar q_{j}$ fermions) and is a light
pseudo-Goldstone boson of some broken global symmetry acting on $q$ and
$\bar q$. The bound state arises because of confining interaction
determined by the running gauge coupling constant $\alpha(Q^{2})$. At one
loop, it is expressed by a familiar formula,
\begin{equation}
\alpha^{\rm 1-loop}(Q^{2}) = 1/\left(4\pi b_{0} \log\left(Q^{2}/\Lambda^{2}
\right) \right),
\label{Eq:10*}
\end{equation}
where
\begin{equation}
b_{0}=\frac{1}{48 \pi^{2}} \left(11 C_{2}^{A} - 4 \sum_{f} T_{f}
\right),
\label{Eq:beta0}
\end{equation}
$C_{2}^{A}$ is the Casimir invariant of the adjoint
representation of the gauge group and $T_{f}/2$ is the Dynkin index for
the representation of fermionic fields
(for $SU(N_{c})$ gauge theory with $N_{f}$ flavours of
fundamental fermions, $b_{0}=\left(11 N_{c} - 4 N_{f}
\right)/(48\pi^{2})$). This introduces the one-loop dynamical scale
$\Lambda$ which, by the definition (\ref{Eq:10*}), is in one-to-one
correspondence with $\alpha^{\rm 1-loop}$ calculated at a certain value of
$Q^{2}$.

To speak about electromagnetic properties of $\Pi$, we allow $q$ and $\bar
q$ to be charged under an extra $U(1)$ gauge group, e.g.\ the
electromagnetic one. Then, the electromagnetic form factor
$F_{\Pi}(Q^{2})$ may be defined in the usual way. It is non-zero in two
cases: either the sum of charges of $q$ and $\bar q$ is nonzero, or it is
zero but $q$ and $\bar q$ have different masses (corresponding QCD
examples are the charged pion and the neutral kaon). In the $\pi$-meson
model, we kept the masses of constituent $u$ ($\bar u$) and $d$ ($\bar d$)
quarks identical, and therefore we keep them identical for $q$ and $\bar
q$ here, though the method can be easily generalized to the other case.

\begin{figure}[tbp]
\centering
\includegraphics[width=\textwidth%
]{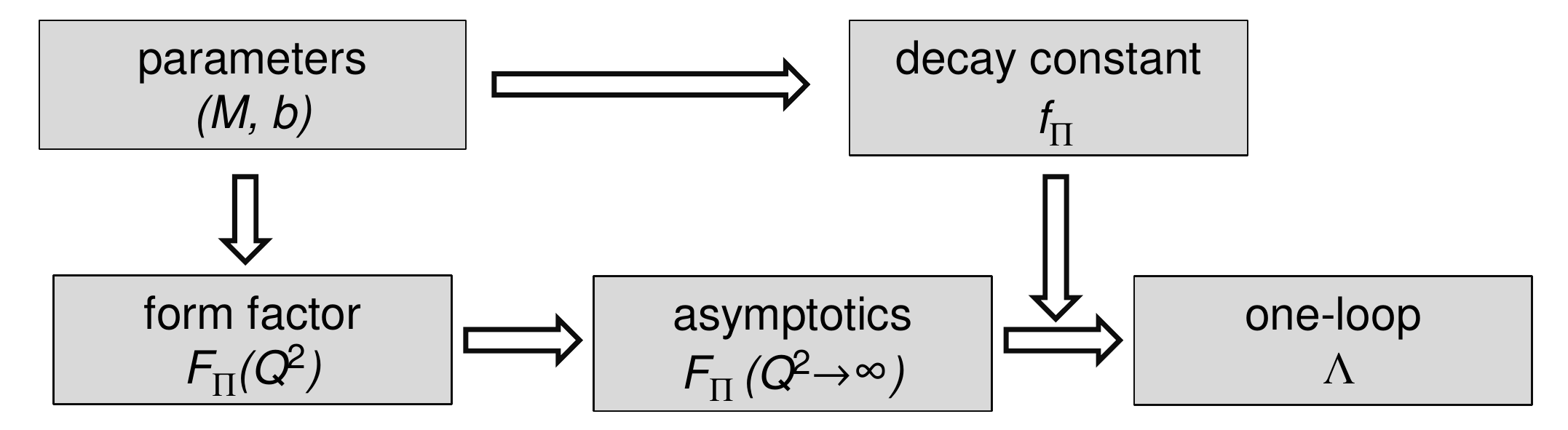}
\caption{\label{fig:sketch}
A scheme of relating phenomenological parameters $(M, b)$ to physical
parameters $(f_{\Pi}, \Lambda)$; see text. }
\end{figure}

The approach described in detail in Refs.~\cite{KrTr-EurPhysJ2001-bundles,
KrTr-PRC2009} (see the Appendix of Ref.~\cite{TrTr-PRD2013-asympt} for a
collection of all necessary formulae) allows one to calculate
$F_{\Pi}(Q^{2})$, starting from two phenomenological parameters, $M$ and
$b$, discussed above. It includes switching off the constituent-quark mass
$M$ smoothly at a certain energy scale, much lower than the values of
$Q^{2}$ at which one expects the high-energy asymptotics to settle down.
An explicit expression for the meson decay constant $f_{\Pi}$ relates it
to the parameters of the model, $M$ and $b$,
\begin{equation}
f_\Pi = \frac{M\,\sqrt{3}}{\pi}\,\int\,\frac{k^2\,dk}{(k^2 +
M^2)^{3/4}}\, u(k) .
\label{Eq:11+}
\end{equation}
On the other hand, at large $Q^{2}$, the gauge-theory
asymptotics~\cite{FarrarJackson, EfremovRadyushkinAs, LepageBrodskyAs} is
\begin{equation}
\left. Q^{2} F_{\Pi} (Q^{2}) \right|_{Q^{2}\to \infty} \to 8\pi
f_{\Pi}^{2}  \alpha^{\rm 1-loop}(Q^{2}).
\label{Eq:11*}
\end{equation}
The right-hand side of Eq.~(\ref{Eq:11*}) depends on two quantities,
$f_{\Pi}$ and $\Lambda$. Since $f_{\Pi}$ is determined from $M$ and $b$ by
means of Eq.~(\ref{Eq:11+}), calculation of $F_{\Pi}(Q^{2})$ at large
$Q^{2}$ allows to determine one-loop $\Lambda$ of the underlying gauge
theory. In this way, we relate two phenomenological parameters, $M$ and
$b$, to two physical parameters, $f_{\Pi}$ and $\Lambda$. Our method thus
makes it possible to calculate $F_{\Pi}(Q^{2})$ in the infrared region,
$Q^{2}<\Lambda^{2}$, starting from $\Lambda$ and $f_{\Pi}$. The logic of
the method is illustrated schematically in Figure~\ref{fig:sketch}.

We turn now to a numerical realization of the method. We illustrate it for
an $SU(2)$ gauge theory with $N_{f}=2$ fundamental fermions. Our approach
does not allow to trace explicitly the influence of the current quark
masses; together with the choice of the gauge group and the matter
content, it determines the value of $f_{\Pi}$ which is treated as
independent parameter of the model (different values of $f_{\Pi}$ at a
fixed value of $\Lambda$ correspond to different gauge groups and/or
different current quark masses). We choose opposite charges of $q_{1}$
($+1/2$) and $q_{2}$ ($-1/2$), so that $\Pi=q_{1} \bar{q_{2}}$ has the
unit charge. As we have already pointed out, we assume equal masses of
$q_{1}$ and $q_{2}$ in this example.

\begin{figure}[tbp]
\centering
\includegraphics[width=0.85\textwidth%
]{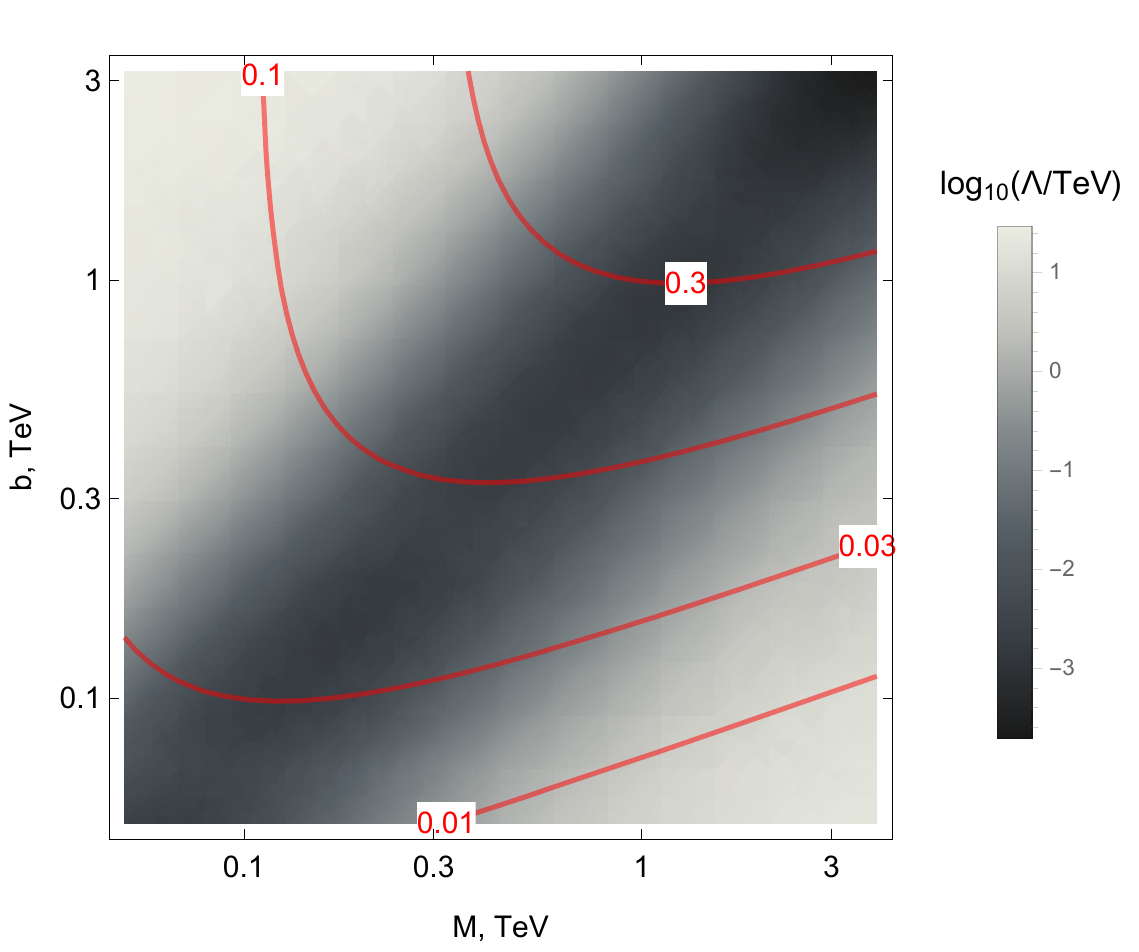}
\caption{\label{fig:plot}
Relation between $(M,b)$ and $(\Lambda, f_{\Pi})$ for an $SU(2)$ gauge
theory with $N_{f}=2$ fundamental fermions. Different levels of shading
correspond to different values of $\Lambda$, see the scale to the right.
Full thick red lines are contours of equal $F_{\Pi}$ corresponding to the
values written on the contours (in TeV).}
\end{figure}

Figure~\ref{fig:plot}
illustrates the calculation outlined above: $f_{\Pi}$ and $\Lambda$
versus $M$ and $b$. All these quantities are dimensionful; keeping in mind
future applications to composite models, see also Sec.~\ref{sec:lat}, for
which $f_\Pi =v =246$~GeV, we change $M$ and $b$
between $0.05$ and $10$~TeV and determine $\Lambda$ from asymptotics above
$Q^{2}=900$~TeV$^{2}$. It is interesting to note that the inverse relation
is not single-valued: as one can see from Figure~\ref{fig:plot}, there are
two pairs of $(M,b)$ corresponding to the same values of
$(f_{\Pi},\Lambda)$ and the branches $M<b$ and $M>b$.

To characterise the behaviour of $F_{\Pi}(Q^{2})$ at $Q^{2}\to 0$, it is
convenient to determine the $\Pi$ charge radius,
$r_{\Pi} \equiv \langle r_{\Pi}^{2} \rangle ^{1/2}$,
\[
r_{\Pi}^{2}=-6 \left. \frac{dF_{\Pi}(Q^{2})}{d Q^{2}} \right|_{Q^{2}\to 0}.
\]
Note that it is the charge radius which determines the cross section
important for the search of composite dark matter states~\cite{0812.3406,
Sannino13}, though they may be more complicated than $\Pi$. While our
approach generates the full $F_{\Pi}(Q^{2})$ function, we will concentrate
on $r_{\Pi}$ for the moment.

\begin{figure}[tbp]
\centering
\includegraphics[width=0.7\textwidth%
]{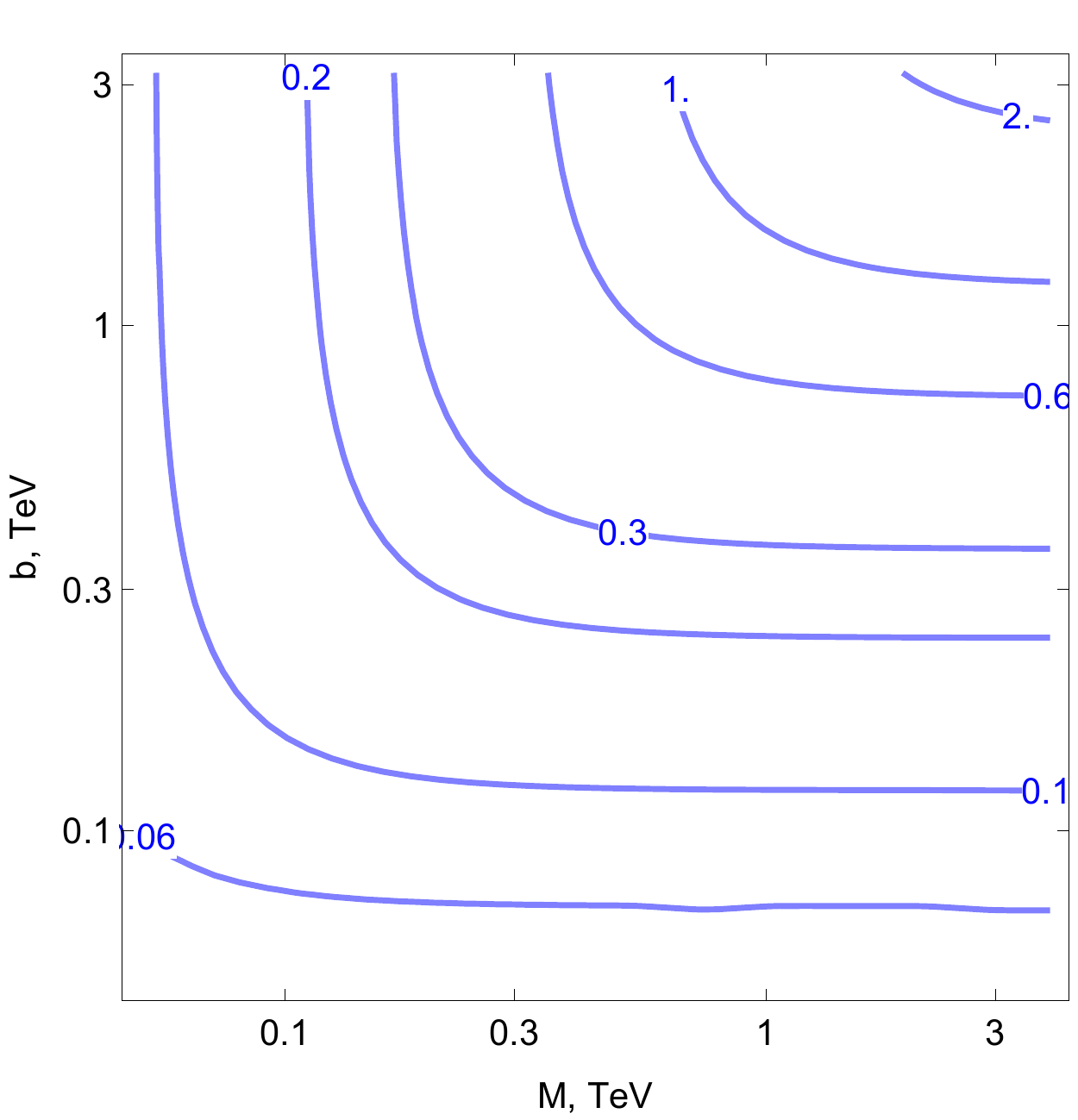}
\caption{\label{fig:rplot}
Inverse charge radius of $\Pi$ versus $M$ and $b$
in an $SU(2)$ gauge
theory with $N_{f}=2$ fundamental fermions.
Full
thick blue lines are contours of equal $1/r_{\Pi}$ corresponding
to the values written on the contours (in TeV).}
\end{figure}

Qualitatively, the behaviour of $r_{\Pi}$ at the two branches, see
Figure~\ref{fig:rplot}, is easily understood. For $M<b$, the theory is in
its strong-coupling regime, the ``quarks'' are light compared to $\Lambda$
and the meson wave-function size in the momentum space, $b$, depends
mostly on $\Lambda$ (the strong interaction) while the size of the meson
$r_{\Pi}$ gets contribution both from $M$ and $b$. The $M>b$ case
corresponds to (relatively) heavy, almost point-like ``quarks'', and the
size of the meson is fully determined by the interaction.

\begin{figure}[tbp]
\centering
\includegraphics[width=0.76\textwidth%
]{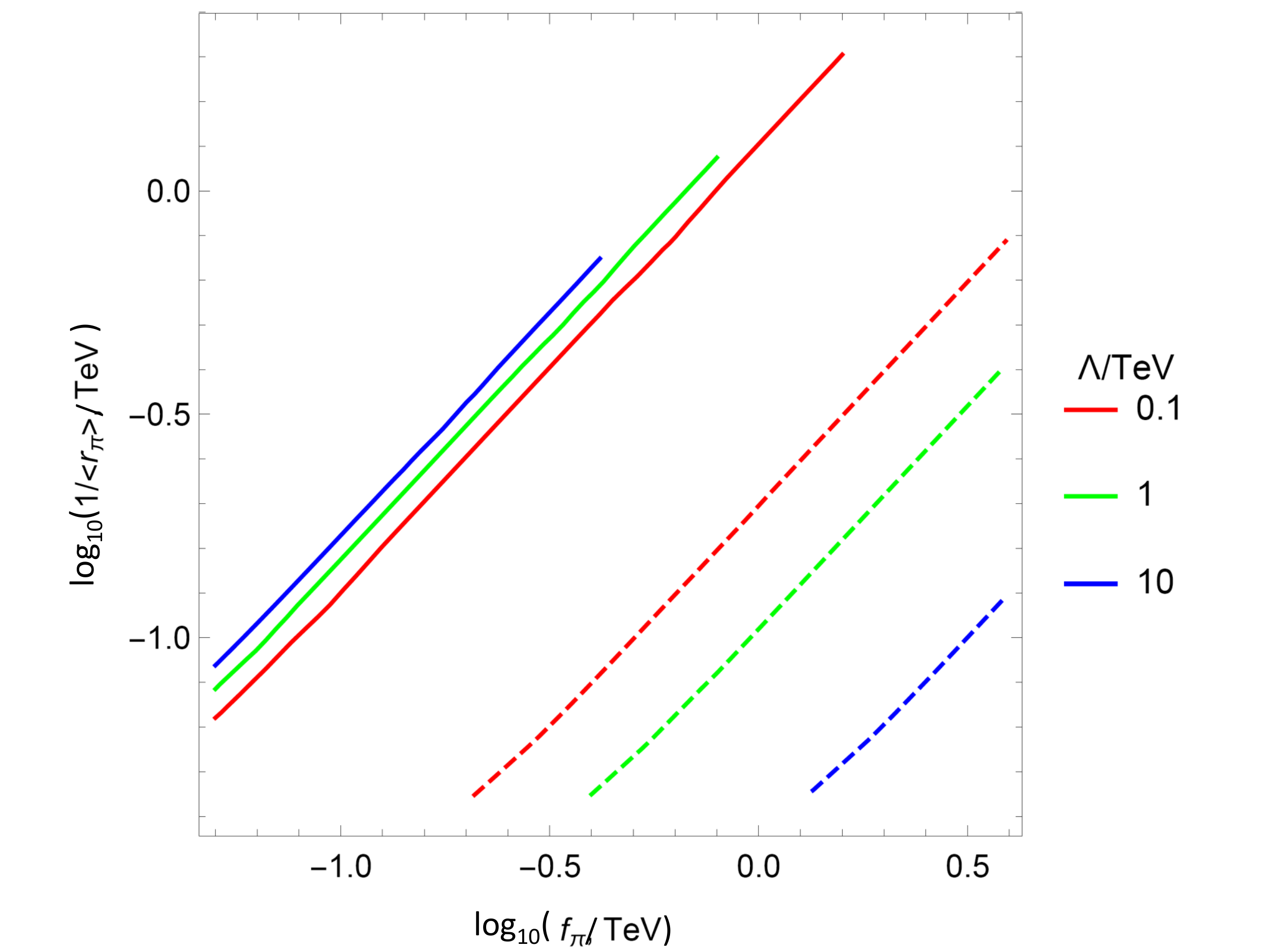}
\caption{\label{fig:rPi}
Inverse charge radius of $\Pi$ versus $f_{\Pi}$
in an $SU(2)$ gauge
theory with $N_{f}=2$ fundamental fermions.
Different colors
correspond to different values of $\Lambda$, see the color code to the
right. Full lines correspond to the branch $M<b$ (``strong coupling''),
dashed lines correspond to the branch $M>b$ (``heavy quarks'').
}
\end{figure}

Figure~\ref{fig:rPi}
presents the dependence of $r_{\Pi}$ on $f_{\Pi}$ for different values of
$\Lambda$. The leading dependence is $1/r_{\Pi} \propto f_{\Pi}$ with the
coefficient depending on $\Lambda$. For the ``strong-coupling'' branch and
for a fixed $f_{\Pi}$, varying $\Lambda$ gives only small (but measurable)
corrections\footnote{For the QCD $\pi$ meson, described by this branch, it
was this correction which allowed to trace the correct $\Lambda$-dependent
numerical coefficient in the asymptotics, see
Ref.~\cite{TrTr-PRD2013-asympt}.}.

\section{Comparison to lattice results}
\label{sec:lat}
In this section, we use published lattice results on $F_{\Pi}(Q^{2})$ in
$SU(2)$ gauge theory with $N_{f}=2$. This result was presented in
Ref.~\cite{Sannino13}, while some required information about the lattice
calculation was given in Ref.~\cite{Sannino14}, based on the same
numerical simulations.

Table~I of Ref.~\cite{Sannino13} gives the values of $F_{\Pi}(\hat Q^{2})$
calculated for three versions of the lattice calculations with different
parameters of the lattice Lagrangian. Here, $\hat Q^{2} \equiv \left( Qa
\right)^{2}$, where $a$ is the lattice spacing, fixed from the condition
$f_{\Pi}=246$~GeV (the latter choice is of course arbitrary and is
motivated by studies of theories with a composite Brout-Englert-Higgs
scalar),
\[
f_{\Pi}=\frac{1}{a}Z_{a}f_{\Pi}^{\rm lat},
\]
$Z_{a}$ is the renormalization constant, whose value used in
Refs.~\cite{Sannino13, Sannino14} is
\[
Z_a=1-k\left(\frac{g_{0}}{4\pi} \right)^{2}
\frac{N_c^2-1}{2N_{c}}.
\]
In this expression, $g_{0}$ is the bare gauge coupling constant (related
to the lattice coupling constant, $\beta$, by $\beta=2N_{c}/g_{0}^{2}$),
$N_{c}=2$ is the ``number of colours'' in the gauge group $SU(N_{c})$ and
$k \approx 15.7$ is a numerical coefficient determined in
Ref.~\cite{0802.0891}. Values of $f_{\Pi}^{\rm lat}$ for all lattice
calculations used are given in Table~II of Ref.~\cite{Sannino14}.

In this way, we know $f_{\Pi}$ and $F_{\Pi}(Q^{2})$, but we also need to
know the gauge coupling constant in the large-$Q^{2}$ limit, or
equivalently one-loop $\Lambda$, to perform a parameter-free test of the
results obtained in our approach. It is, generally, a nontrivial task to
relate $\Lambda$ and $g_{0}$, because lattice and continuum models use
different ways of renormalization (see e.g.\
Refs.~\cite{9711243, 1004.3462} for reviews). To extract the physical value
of $\alpha$ or $\Lambda$ from a lattice calculation, a certain observable
(related, for instance, to the force of interaction between fermions) is
usually calculated. These calculations have not been performed for
configurations used in Ref.~\cite{Sannino13}.

Fortunately, for our purposes, it is sufficient to follow a different
approach. Indeed, what we need in
Eq.~(\ref{Eq:11*}) is, by definition, the one-loop coupling constant which
enters the asymptotics at large $Q^{2}$. In the asymptotical region, one
can expand coupling constants, determined in different schemes, in powers
of each other (see e.g.\ Ref.~\cite{hep-lat/0211036}). One obtains an
expression for the lattice strong-coupling scale $\Lambda_{\rm lat}$,
\[
\Lambda_{\rm lat}=\frac{1}{a} \left(b_{0} g_{0}^{2}
\right)^{-\frac{b_{1}}{2b_{0}^{2}}} \exp \left( -\frac{1}{2 b_{0}g_{0}^{2}}
\right)
\times c,
\]
where the correction
\[
c=\exp \left[-\int\limits_{0}^{g_{0}} \! \left( \frac{1}{\beta(t)} +
\frac{1}{b_{0} t^{3}} - \frac{b_{1}}{b_{0}^{2}t} \right)\,dt \right]
\simeq
1+ \mathcal{O} (g_{0}^{2})
\]
and the two-loop beta function is
\[
\beta(t)=-t^{3} \left(b_{0}+ b_{1} t   \right);
\]
the coefficient $b_{0}$ is determined in Eq.~(\ref{Eq:beta0}) and, for an
$SU(N_{c})$ gauge theory with $N_{f}$ flavours of fermions in the
fundamental representation,
\[
b_{1}= \frac{1}{(4\pi)^{2}} \left( \frac{34}{3} N_{c}^{2} - \frac{10}{3}
N_{c}N_{f} - \frac{N_{c}^{2}-1}{N_{c}}N_{f} \right).
\]
The relation between the continuum $\Lambda$ (determined, e.g., in the
${\overline{\rm MS}}$ renormalization scheme) and $\Lambda_{\rm lat}$ reads
as
\[
\Lambda= \Lambda_{\rm lat} \exp \left(-\frac{l_{0}}{2 b_{0}}  \right),
\]
where
\[
l_{0}= \frac{1}{8 N_{c}}+k_{c}N_{c} + k_{f}N_{f},
\]
the coefficient $k_{c}=-0.16995599$ was determined in
Ref.~\cite{0709.4368} and, for the Wilson fermions used in the lattice
calculation we discuss, $k_{f}=0.0066959993$, see Ref.~\cite{0602023}.

The precision of this method is not very high. The main source of errors
is in the use of perturbative expressions for $\Lambda$ and $Z_{a}$ (the
latter affects $\Lambda_{\rm lat}$ through the value of $a$). Particular
values of the coupling constants used for the simulations we address
correspond to $g_{0}\sim 1.4$, so that the precision is limited by the
loop factor $\sim g_{0}^2/(4\pi)\sim 16 \%$. Additional uncertainties
appear in the lattice calculation of $f_{\Pi}^{\rm lat}$ and of other
quantities, so we use a conservative estimate of $\pm 20\%$ precision in
$\Lambda$ (remember that $f_{\Pi}$ is fixed and all uncertainties in
$f_{\Pi}^{\rm lat}$ are translated into those of $a$).

Table~\ref{tab:i}
\begin{table}[tbp]
\centering
\begin{tabular}{|c|ccccccc|}
\hline
Case&$(\beta, m_{0})$& $g_0$ & $Z_{a}$ & $f_{\Pi}^{\rm lat}$ & $1/a$,
GeV & $\Lambda_{\rm lat}$, GeV & $\Lambda$, GeV\\
\hline
I & (2.2,$-$0.72) & 1.348 & 0.864 & 0.0664(5) & 4286 & 9.04 & 298\\
II& (2.2,$-$0.75) & 1.348 & 0.864 & 0.0457(7) & 6227 & 13.14& 423\\
III&(2.0,$-$0.947)& 1.414 & 0.851 & 0.0754(8) & 3834 & 15.03& 497\\
\hline
\end{tabular}
\caption{\label{tab:i} Parameters of the lattice calculations (see text).}
\end{table}
lists three different sets of lattice data we use, determined by $\beta$
and by the bare fermion mass $m_{0}$, together with useful values of
corresponding parameters. Table~\ref{tab:ii}
\begin{table}[tbp]
\centering
\begin{tabular}{|c|ccc|}
\hline
Case& \multicolumn{3}{|c|}{$r_{\Pi}$, TeV$^{-1}$} \\
&lattice & $M<b$ & $M>b$\\
\hline
I & 1.6 & 1.9 & 1.0\\
II& 1.2 & 1.9 & 1.0\\
III&1.4 & 1.8 & 1.0\\
\hline
\end{tabular}
\caption{\label{tab:ii} Comparison of $r_{\Pi}$ obtained in our
calculation with lattice results.}
\end{table}
gives, for the three
cases, results of the lattice calculations of $F_{\Pi}(Q^{2})$ (data
points) together with calculations by our method,
Section~\ref{sec:main}, for values of $f_{\Pi}=246$~GeV and $\Lambda$
given in Table~\ref{tab:i}.
We see that, within the precision behind these
numbers, the agreement is reasonable, which represents a highly nontrivial
test of our method.

\section{Conclusions and outlook}
\label{sec:concl}
In this work, we started from a method to calculate the electromagnetic
form factor of the $\pi$ meson, $F_{\pi}(Q^{2})$, which has been shown
previously (i)~to predict experimental values of $F_{\pi}(Q^{2})$,
measured later, without tuning of parameters, and (ii)~to obtain correct
QCD asymptotics at $Q^{2}\to \infty$, again with no additional parameters
introduced or tuned. This allowed us to link infrared and ultraviolet
regimes in a non-trivial way. Then we pretended that this method is
general and applied it to non-QCD gauge theories, concentrating on a
pion-like state $\Pi$. We presented a method to obtain general numerical
relations between intrinsic parameters of the theory, $f_{\Pi}$ and
$\Lambda$, and phenomenological parameters of the model, which allows one
to calculate the form factor, $F_{\Pi}(Q^{2})$, starting from known values
of these physical parameters, without fitting anything to experimental
data. We then tested the outcome of this method versus known lattice
results for $F_{\Pi}(Q^{2})$ for a $SU(2)$ gauge theory with $N_{f}=2$
flavours of fundamental fermions and obtained a reasonable agreement in
accordance with the precision, limited mostly by uncertainties in
determination of $\Lambda$ corresponding to the lattice calculations.

Together with the success in the description of the
real QCD $\pi$ meson, this result supports the proposal that our method,
based on Relativistic Hamiltonian Dynamics and presumably aimed at low
energies only, may nevertheless be used for calculation of electromagnetic
properties of bound states in strongly coupled gauge theories defined in
their ultraviolet limit. The physical reason behind this is probably
related to the relativistic invariance carefully preserved throughout our
calculations, in contrast with some other approaches.

At least in its present form, the approach is not universal since (i)~it
addresses a particular problem, calculation of the electromagnetic form
factor, and (ii)~it uses, as the input parameters, the gauge coupling
constant (through $\Lambda$) and the decay constant of the pion-like
state, $f_{\Pi}$. The latter, in principle, should be expressed through
parameters of the Lagrangian, the coupling constant and fermion masses,
but this is beyond the capabilities of our method. Still, it is an unusual
success and opens the possibility of physical applications, the most
straightforward one dealing with composite-Higgs models with composite
dark-matter particles. In this case, $f_{\Pi}$ is fixed from the
electroweak symmetry breaking and is therefore known. The dark-matter
particle is another, different from $\Pi$, meson, but its
phenomenologically interesting cross section is determined by its
electromagnetic form factor~\cite{0812.3406}, which can be calculated by
the method suggested here. This interesting approach will be followed
elsewhere.

\begin{acknowledgments}
We are indebted to Victor Braguta, Sergei Dubovsky, Dmitry Levkov and Yury
Makeenko for interesting and helpful discussions. We thank Andrei Kataev
for pointing out a misprint in the definition of the beta function. The
work of ST on nonperturbative description of strongly coupled bound states
is supported by the Russian Science Foundation, grant 14-22-00161.
\end{acknowledgments}

\end{document}